**Title**

**Hong-Ou-Mandel interferometry and spectroscopy using entangled photons**

**Authors**

Konstantin E. Dorfman,[1]* Shahaf Asban,[2] Bing Gu[2], and Shaul Mukamel[2]

1. State Key Laboratory of Precision Spectroscopy, East China Normal University, Shanghai 200062, China

*Email: dorfmank@lps.ecnu.edu.cn, ORCID: 0000-0001-9963-0878

2. Department of Chemistry and Physics & Astronomy, University of California, Irvine, California 92697-2025, USA

**Abstract**

Optical interferometry has been a long-standing setup for characterization of quantum states of light. Both the linear and the nonlinear interferences can provide information about the light statistics an underlying detail of the light-matter interactions. Here we demonstrate how interferometric detection of nonlinear spectroscopic signals may be used to improve the measurement accuracy of matter susceptibilities. Light-matter interactions change the photon statistics of quantum light, which are encoded in the field correlation functions. Application is made to the Hong-Ou-Mandel two-photon interferometer that reveals entanglement-enhanced resolution that can be achieved with existing optical technology.

**Introduction**

Quantum states of light provide an exciting platform for observing and controlling matter beyond what is possible classically [1-4]. Quantum states are very sensitive to the external environment which makes them useful probes of matter. Quantum features of light have long been used in metrology and quantum information while lately there has been a growing activity in utilizing them in spectroscopic applications. Interferometry offers a robust detection schemes of quantum light. In this paper we present a novel spectroscopy based on the Hong-Ou-Mandel (HOM) [5] two-photon interferometric setup. Observables measured by the interference of two waves depend on two times separated by a delay $\Delta$ which can be controlled by the propagation path difference of the mixed waves. The unified picture of second-order and fourth-order interferences in a single interferometer have been demonstrated in [6]. Previous works attempted either to utilize HOM-like measurements to address properties of the beam splitters or the individual emitters [7] or utilize four-wave mixing for the quantum light generation [8] and characterization [9].

Interferometric signals can be recast in terms of moments of the field operators posterior to the interaction with the matter, thereby revealing its statistics. The signal- field operator, (see Methods and Supplementary Note 1) is given by [10]

$$\hat{\mathbf{E}}_s^{(+)}(\mathbf{r},\mathbf{t}) = -i\omega_s \hat{V}(\mathbf{r},t), \quad (1)$$

Where $\omega_s$ is the signal mode frequency and $\hat{V}$ is the matter dipole operator. The interferometric setup naturally gives rise to two characteristic timescales and respective length scales. First, the response interval $\tau_R$ in which light-matter interaction occurs. It is determined by the pulse envelope, the spatial dimension of the sample and the response time. Second is the pulse relative delay interval $\delta\tau$ determined by the interference region governed by the interferometer dimensions. Here, we consider the field-matter interaction region to be localized compared to the spatial dimensions of the interferometer. We further consider a sequence of ultrafast coherent excitation pulses – which are classical for all practical purposes – followed by an interaction with the quantum state of light. The interferometer operation mode is shown in Fig. 1: when the pulse interval $c\delta T$ and the maximal response interval of the sample $c\tau_R$, are smaller than the free propagation distance between the sample and the interference-detection location $L_p$ $(L^* \ll L_p)$ where $L^* \coloneqq c\max\{\tau_R, \delta T\}$ - the measured response functions are classical. The response in this regime is highly localized temporally and immediately after the pulse interacts with the sample, the matter degrees of freedom can be traced out. The excitation and deexcitation period is dominated by the duration of the narrowband envelope of the quantum field given by $c\tau_R$ combined with the pulse delay interval $c\delta\tau$. The maximal duration of this interval is defined by $L^*$ which is smaller than a few hundred micrometers even for a picosecond pulse which is well within the narrowband region. For example, for a transform limited Gaussian pulse with central wavelength $\lambda_c = 1064\ nm$ and pulse duration $\Delta t = 2\ ps$ which occupies $c\tau_R \approx 600\ \mu m$. Finally, the interferometer length scale denoted $L_p$ specifies the free propagation distance between the incident beams, the beam-splitter and the detector. Typical interferometer length is in the order of few centimeters which justifies the separation of timescales – considering the interaction interval to be localized around the sample compared with $L_p$. Moreover, the Rayleigh distance is typically a few meters in this setup, thus one can consider the propagation as unidirectional for all practical purposes. For a femtosecond pulse $c\delta T \propto 10^{-1} \mu m$ while for a traditional interferometer $L_p$ is in the order of centimeters. In the following we consider short pulses, so that $(L^* \ll L_p)$. Trace w.r.t. matter results in the $n^{th}$ order polarization in the external field $P^{(n)}(t) \equiv \langle \hat{V}(t) \rangle_{\{n\}}$ (details are given below). This polarization serves as a source for the signal field. This regime fits experimental setups involving ultrafast pulses ($\delta T \leq 1\, ps$). In the opposite regime, $c\tau_R, c\delta T \gg L_p$, the pulse is long enough to create ambiguity in the order of interactions and the arrival of the relatively delayed photons (see Supplementary Notes 2-4 for detailed derivations of this regime). One cannot then trace the matter degrees of freedom prior to the measurement which gives rise to different observables.

In the present work we combined the interferometric detection (HOM) with wave mixing that involves both classical and quantum light beams to address more complex nonlinear optical processes and the corresponding components of the nonclassical response function. We investigate how the quantum state of light and its statistics are modified by interaction with matter. In particular, we address the following two issues of the quantum nonlinear interferometric spectroscopy. The first issue is regarding the nature of the change of the quantum state and its statistics. The second point investigates the details of the matter information that can be deduced from the change in the statistics of the field. These questions are explored by using an interferometric setup traditionally used to study quantum states of light and now applied for investigation of the matter degrees of freedom via extraction of the matter response functions. We therefore focus on accuracy enhancement of such responses and their deviations from classical susceptibilities.

**Results and Discussion**

The proposed experiment combines several conventionally used optical techniques such as four-wave mixing, beam splitting, and Hong-Ou-Mandel (HOM) interferometer, and three-photon absorption spectroscopy. In the following sections we present each technique independently highlighting the main principle and the underlying theoretical model that will be used to describe each part of the setup. We finally combine both techniques in the setup shown in Fig. 2a and discuss the resulting HOM spectroscopic measurements. In the experimental setup the three classical light beams generated are combined with the quantum light pulse produced by the parametric down conversion (independently from the classical three pulses). The corresponding level -scheme is shown in Fig. 2b. The main goal of the proposed measurement is to use interferometric (HOM-like) detection to investigate the $\chi^{(3)}$ nonlinear susceptibility that combines an absorption of the three classical fields and the transmission of one quantum field shown in Fig. 2c. The corresponding Feynman diagram and HOM signal are shown in Fig. 2d and e, respectively. Unlike the Kerr process which requires high intensity laser pulses to produce third-order nonlinear response, here we deal with resonant absorption of each of the light beams participating in the four-wave mixing. This nonlinear process is the main focus of our study.

The general third order nonlinear optical process generates various signals that are well studied in classical light spectroscopy: optical pump-probe, Raman, fluorescence, transient grating, photon echo and others. Four-wave mixing (FWM) signals plays an important role as it allows to have additional control over the field-matter interactions via spatial phase matching. The typical FWM setup shown in Fig. 3 where three beams interacting with the material sample generate a fourth beam propagating in the direction governed by one of the eight possible phase matching conditions. The response functions are obtained by tracing over the matter degrees of freedom. The state of the outgoing field in Fig. 3 is given by tracing Eq. (1) over the matter degrees of freedom $\hat{\mathbf{E}}_a^{(+)}(t) \rightarrow \hat{\mathbf{E}}_s^{(+)}(t)$,

$$\hat{\mathbf{E}}_s^{(+)}(\mathbf{r},t) = -i\omega_s \sqrt{\frac{\hbar\omega_s}{2\pi}} Nf(\Delta k) \int d\omega \chi^{(3)}(\omega) \hat{a}(\omega) e^{-i\omega t}, (2)$$

where $\chi^{(3)}(\omega) \equiv \chi^{(3)}(-\omega; \omega_3, \omega_2, \omega_1)$ is the third-order susceptibility. We have omitted the three classical incoming wave frequencies for brevity. The matter is modeled by a collection of N homogeneously distributed point-like molecular dipoles at random positions $\mathbf{r}_\alpha$. Adopting the multipolar coupling Hamiltonian following with rotational averaging we obtain

$$\langle \hat{V}(r,t) \rangle_{\{n\}} = \sum_{\alpha=1}^{N} \delta(\mathbf{r} - \mathbf{r}_\alpha) \langle \hat{V}(t) \rangle_{\{n\}} e^{i\mathbf{k}_{\{n\}}\mathbf{r}}. (3)$$

Here $\{n\}$ denotes averaging with respect to the $n^{th}$ order density operator due to $n$ interacting fields one of which is the photon with the entangled noninteracting counterpart. In this calculation each of the incoming modes interacts with a single molecule. $f(\Delta k) = \frac{1}{N}\sum_\alpha e^{i\Delta k \mathbf{r}_\alpha}$ is a geometrical factor that carries the information regarding the distribution of molecules which gives rise to the phase matching condition when $\Delta k = 0 = \mathbf{k}_{\{n\}} - \mathbf{k}_s$ and $\mathbf{k}_{\{n\}} = \sum_{i=1}^{3} \pm \mathbf{k}_i$ where $\mathbf{k}_s$ is the wavevector of the entangled photon containing the $2^3$ phase matching directions [11]. Note that $f(0) = 1$.

**The Hong-Ou-Mandel interferometric signal.** We next turn to the HOM two-photon interferometer in the presence of matter. The electric field is transformed by the relatively displaced beam-splitter (BS) depicted in Fig. 4 according to

$$\begin{pmatrix} \hat{\mathbf{E}}_{a\prime}^{(+)}(\omega) \\ \hat{\mathbf{E}}_{b\prime}^{(+)}(\omega) \end{pmatrix} = \begin{pmatrix} \sqrt{T} & i\sqrt{R}e^{i\omega\Delta} \\ i\sqrt{R}e^{-i\omega\Delta} & \sqrt{T} \end{pmatrix} \begin{pmatrix} \hat{\mathbf{E}}_{a}^{(+)}(\omega) \\ \hat{\mathbf{E}}_{b}^{(+)}(\omega) \end{pmatrix}, (4)$$

where the linear phase results in the $\pm\Delta$ relative time delay, corresponding to the $\pm c\Delta$ displacement of the BS. $\sqrt{T}$ and $\sqrt{R}$ are the transmission and reflection coefficients. We focus on the photon coincidence signal depicted in Fig. 1 given by a joint probability to detect one photon in $D_a$ and one photon in $D_b$ separated by delay $\tau$ given by

$$\mathcal{N} P_{ab}(\tau) = \langle \mathrm{T} \hat{E}_{a\prime,R}^{(-)}(t) \hat{E}_{b\prime,R}^{(-)}(t+\tau) \hat{E}_{b\prime,L}^{(+)}(t+\tau) \hat{E}_{a\prime,L}^{(+)}(t) \rangle, (5)$$

where $\mathcal{N}$ is a normalization factor. We employ the superoperator notation, $O_L A = OA$ and $O_R A = AO$, the superoperator $O_\pm$ represents an anti/commutator $O_\pm A = OA \pm AO$. Note that the superoperator time ordering T, which is an operator in Liouville space is different from the standard Glauber's normally ordered operators [19]. The plus-minus and the left-right superoperators are linked by a linear transformation. Below we focus on a narrowband pump. Extension to a broadband pump is outlined in Supplementary Note 2.

**The narrowband HOM spectrometer.** In their seminal paper, HOM have used the narrowband wavefunction (see Eq. (16) and Methods). Following this procedure where in path $'a'$ (top branch of the interferometer) a sample composed of many molecules is placed, and the signal is given by the four-wave mixing setup depicted in Fig. 3.

We focus on the $L^* \ll L_p$ regime, where the spatial extent of the photon wavepacket after the interaction is small compared to the dimensions of the interferometer. Calculating the coincidence count according to Eq. (5) using Eqs. (2) and (15) we obtain,

$$P_{ab}(\tau, \Delta; \{\omega\}_n) = P_0 \{ T^2 |C(\tau)|^2 + R^2 |C(2\Delta - \tau)|^2 - RT[C^*(\tau)C(2\Delta - \tau)e^{-i\omega_p(\tau-\Delta)} + c.c.] \}.$$

(6)

Here the convoluted response is given by the functional $C(\tau) = G(\tau) * \chi^{(3)}(\tau)$, where $G(\tau) = (2\pi)^{-\frac{1}{2}} \int d\omega \, \Phi(\omega, \omega_p - \omega) e^{-i\omega\tau}$, where the two-photon wavefunction amplitude $\Phi(\omega_a, \omega_b)$ is given by Eq. (15) (see Methods); $\chi^{(3)}(\tau) = (2\pi)^{-\frac{1}{2}} \int d\omega \, \chi^{(3)}(\omega) e^{-i\omega\tau}$ and $\{\omega\}_3 = \omega_1, \omega_2, \omega_3$ are the frequencies of the three classical waves. The pre-factor containing the central frequency and the setup geometry is given by $P_0 = \mathcal{N}^{-1/2}(N\hbar\omega_0^2)^2 |f(\Delta k)|^2$. In the absence of matter, $\chi(t) = \delta(t)$ and we recover the HOM interference [5] signal. In that case the extra phase factor that appears in the second term in Eq (6) can be shifted at the frequency integration by $\omega \to \omega + \omega_p/2$. When the material sample is added, a reference frequency is set. This can be compensated by equivalently translated matter response. When the coincidence counting is not temporally gated, we obtain the signal by integration over $\tau$,

$$P_{ab}(\Delta, \{\omega\}_n) = n_0 - v \int d\tau \left\{ C^*(\Delta + \tau) C(\Delta - \tau) e^{-i\omega_p \tau} + c.c. \right\}. \quad (7)$$

where $n_0 = P_0(R^2 + T^2) \int d\tau \, |C(\tau)|^2$, and $v = P_0 RT$. For large BS displacement $c\Delta$, the overlap term - the second term in the R.H.S. of Eq. (7) - vanishes due to diminishing correlations of the relatively shifted response. It assumes a Wigner-function form in the $\{\Delta, \omega_p\}$ space. As $\Delta$ is reduced, the overlap term increases, introducing the hallmark dip in the HOM interference pattern. A material sample added in one of the pathways affects the overlap term. Matter information is revealed in Eq. (7) by the variation of the HOM dip with the convoluted response $C(\tau)$ Wigner-function as illustrated in Fig. 2e. Note, that the response function $C(\tau)$ is calculated using three classical fields followed by a single photon field in the last interaction. While it is not unusual that quantum enhanced performance is dramatically eroded by the loss of a single photon, this is not the case here. Several noise sources can be considered such as losses associated with single photon sources, non phase matched contributions, and imperfect transmission and detection efficiency. First, the proposed setup in the single photon regime allows overcoming the noise because of the photon correlation measurement. The classical incoming fields contain a large number of photons and are thus insensitive to losses compared to other classical technique. When the single photon contribution has losses, the signal vanishes due to violation of the phase-matching due to momentum conservation. Second, the improved performance is attributed to the non-classical correlations between the photon pair, not from their Fock-state characteristics. Contributions to such losses can originate from non-phase matched signals, like spontaneous emission adding vacuum fluctuations to the transmitted beams. Third, losses occurring after the mixing can be modeled by a beam splitter with an empty port [27]. Recent interferometry. Experiments in the four-wave mixing setup performed in a multiphoton regime [28] indicate the reduction of quantum correlations is proportional to the square of the transmission ratios of the light beam intensities that characterize such losses. Single photon experiments have substantially lower transmission ratios and are therefore robust against such losses. Finally, it has been shown that similar biphoton spectroscopy measurements are robust against the external noise at the detection stage such as background thermal radiation [29], even under the signal-to-noise ratio reaches 1/30.

**Interferometric detection of $\chi^{(3)}$.** We now turn to the setup depicted in Fig. 2a. The. molecule is modelled by a four level system $\{g, e_1, e_2, f\}$ with transition energies $\omega_{e_1 g} = E_{e_1} - E_g = 3\ eV$, $\omega_{e_2 e_1} = 2\ eV$, $\omega_{f e_2} = 1\ eV$. It undergoes three interactions with classical light pulses with controlled delays, the fourth-interaction is taken to be one photon produced by a parametric down-converted shown in Fig. 2a. The signal with phase matching at $\boldsymbol{k}_s = \boldsymbol{k}_a + \boldsymbol{k}_b + \boldsymbol{k}_c$ can be generated in the four electronic state system. One can surely select another phase matching direction, out of the eight possible ones. Each direction contains a different type of material information governed by the set of pathways containing in the corresponding $\chi^{(3)}$ nonlinear susceptibility. The third order nonlinear susceptibility is calculated using perturbative field-matter interactions arranged in the diagram in Fig. 2d. Following the general approach outlined in chapter six of Ref. [12] we obtain:

$$\chi^{(3)}(-\omega; \omega_c, \omega_b, \omega_a) = \frac{-\mu_{f e_2} \mu_{e_2 e_1} \mu_{e_1 g}}{(\omega_a + \omega_b + \omega_c - \omega_{fg} + i\gamma_{fg})(\omega_a + \omega_b - \omega_{e_2 g} + i\gamma_{e_2 g})(\omega_a - \omega_{e_1 g} + i\gamma_{e_1 g})}, \quad (8)$$

where $\omega = \omega_a + \omega_b + \omega_c$ implies energy conservation and $\mu_{ij}$ are transition dipole moments. The nonlinear susceptibility contains one-, two-, and three-photon resonances determined by the transition frequencies $\omega_{ij}$ and dephasing rates (linewidth) $\gamma_{ij}$, $i, j = g, e_1, e_2, f$.

Fig. 2e compares the HOM signal $P_c(\Delta)$ with and without matter. Without matter, the spectrum shows the well-known HOM dip. We consider three classical beams with central frequencies matching the $\omega_{e_1 g}$, $\omega_{e_2 e_1}$ and $\omega_{f e_2}$. Here $\omega_{mn} = E_m - E_n$ is the transition frequency for $|n\rangle \to |m\rangle$. We assume that the matter-induced modulation of the HOM spectrum measures the susceptibility $\chi^{(3)}(-\omega; \omega_3, \omega_2, \omega_1)$ obtained by scanning $\omega_3$ while $\omega_1$ and $\omega_2$ are fixed. The main contribution to $\chi^{(3)}$ is represented by the ladder diagram shown in Fig. 2d. The sample breaks the time-reversal symmetry $P(\tau) = P(-\tau)$ possessed by the bare HOM dip. Such symmetry is a consequence of the exchange symmetry in the twin-photon wavefunction $\Phi(\omega_1, \omega_2) = \Phi(\omega_2, \omega_1)$. By modulating one arm of the twin-photon, the interaction with matter breaks the exchange symmetry. When absorption can be neglected, the matter acts as a frequency-dependent phase shifter $T(\omega) \equiv e^{i\theta(\omega)}$, the optical delay in the idler beam can compensate the modulation if $\Delta$ and $\theta$ have the same sign, or further enhance the relative phase difference between the two beams.

The measurement resolution can be controlled by photon entanglement. Fig. 5 compares the HOM signal using a nonentangled photon pair (a, b) and highly entangled twin photons (c, d). The two-photon wavefunctions is shown in panels a and c, and the corresponding HOM signals are shown in panels b and d. The coincidence HOM measurement can reveal the energies and lifetimes of the four electronic levels $\{g, e_1, e_2, f\}$ along with the transition dipoles between the states and relevant coherence dephasing rates. For instance, the f-g coherence dephasing is a crucial parameter that determines the temporal resolution of the coincidence measurement and strongly depends on the state of light. The signal in Fig. 5 is computed by scanning the pump frequency impinging the nonlinear crystal. As shown, the central feature arises at $\frac{\omega_p}{2} = 6\ eV$ which matches the transition frequency $\omega_{fg}$. The decay of the signal comes from the finite lifetime of the $|f\rangle\langle g|$ coherence induced by the classical laser pulses. While both classical and quantum light give the resonance frequency, the temporal resolution that can track the f-g coherence decay is significantly enhanced by using a highly entangled photon pair. This can be understood in a similar way to the two-photon absorption with entangled photons where the product of temporal and spectral resolutions violates the uncertainty relation [19].

The quantum light statistics is modified by interaction with matter. To understand the nature of the change and the matter information it carries, we have used the HOM interferometric setup which provides information about the state of light after the interaction with matter. Ultrashort pulses can exploit the quantum nature of light in order to increase the measurement accuracy of a classical response function. This was studied in detail for a model system.

Interaction of quantum systems changes their state in the course of light-matter interaction at a single-photon regime [23-25]. Each interaction enhances the correlation, and the system becomes more inseparable. This may be employed in novel quantum spectroscopic setups, which extract matter information from optical probes. While single-photon states can be easily described in the photon number (Fock) basis, they are less suitable for probing phase-shifts due to number-phase uncertainty [13,14]. Multiphoton (entangled) states provide a richer playground for improving the temporal resolution imprinted by matter on the optical probe and is the focus of our study. Multimode squeezed states [15] may be useful since the number-phase uncertainty can be further tuned in order to reach a desired joint frequency-time resolution.

## Methods

**Optical signals description.** We start with the joint light-matter Hamiltonian,

$$\hat{H} = \hat{H}_\mu + \hat{H}_\phi + \hat{H}_{\mu\phi}, \quad (9)$$

where $\mu, \phi$ represent the matter and electromagnetic field respectively, and $\hat{H}_{\mu\phi}$ is their coupling. The electric field operator is partitioned into positive and negative frequency components,

$$\hat{\mathbf{E}}(\mathbf{r},t) = \sum_{s=1}^{n+1} \hat{\mathbf{E}}_s^{(+)}(\mathbf{r},t) + \hat{\mathbf{E}}_s^{(-)}(\mathbf{r},t),$$
$$\hat{\mathbf{E}}_s^{(+)}(\mathbf{r},t) = \left[\hat{\mathbf{E}}_s^{(-)}(\mathbf{r},t)\right]^\dagger, \quad (10)$$

where sum over s runs over the modes participating in the wave-mixing experiment, the positive frequency component (h.c. of the negative counterpart) in the continuum limit is expressed in the slowly varying amplitude approximation,

$$\hat{\mathbf{E}}_s^{(+)}(\mathbf{r},t) = \left(\frac{\hbar\omega_s}{2\pi}\right)^{\frac{1}{2}} e^{i\mathbf{k}_s\mathbf{r}} \int d\omega\, \hat{a}_s(\omega) e^{-i\omega t} \quad (11)$$

with the canonical bosonic commutation relations $[a_s(t), a_{s'}^\dagger(t')] = \delta_{ss'}\delta(t-t')$. Given the dipole operator $\hat{\mu} = \hat{V} + \hat{V}^\dagger$, the light-matter interaction in the Rotating Wave Approximation (RWA) is given by,

$$\hat{H}_{\mu\phi} = \hat{\mathbf{E}}^{(-)}(\mathbf{r},t)\hat{V}(\mathbf{r},t) + \hat{\mathbf{E}}^{(+)}(\mathbf{r},t)\hat{V}^\dagger(\mathbf{r},t), \quad (12)$$

which associates absorption of a photon with dipole excitation and emission with de-excitation. We shall solve the equation of motion of the field in the interaction picture,

$$\frac{d}{dt}\hat{\mathbf{E}}_s^{(+)}(\mathbf{r},\mathbf{t}) = -\frac{i}{\hbar}\left[\hat{\mathbf{E}}_s^{(+)}(\mathbf{r},\mathbf{t}), \hat{H}_{\mu\phi}(\mathbf{r},\mathbf{t})\right]. \quad (13)$$

**The entangled two-photon wavefunction.** We next present the general form of the entangled two-photon wavefunction. Eqs. (15), (16) and (S10) are then used for two limiting cases of narrowband and broadband generating pulses. The wavefunction of frequency-entangled photons, generated by a parametric down-conversion (PDC) is given by [16,17],

$$|\psi\rangle = \int d\omega_a d\omega_b \Phi(\omega_a,\omega_b) a^\dagger(\omega_a) b^\dagger(\omega_b) |0_a, 0_b\rangle, \quad (14)$$

where $\Phi(\omega_a,\omega_b)$ is the two-photon amplitude and $a^\dagger(\omega)$ and $b^\dagger(\omega)$ are creation operators for the two modes. Here $\omega_a/c$ and $\omega_b/c$ correspond to the projections of the wavevectors along the crystal length, where c is a speed of light. These operators obey boson commutation relations

$[a(\omega), a(\omega')] = [b(\omega), b(\omega')] = [a(\omega), b^\dagger(\omega')] = 0$ and $[a(\omega), a^\dagger(\omega')] = [b(\omega), b^\dagger(\omega')] = \delta(\omega - \omega')$. For a type II phase-matched PDC process generated by a broadband pulse, the amplitude is given by pump envelope multiplied by the phase matching factor

$$\Phi(\omega_a, \omega_b) = \alpha(\omega_a, \omega_b)\varphi(\omega_a, \omega_b). \quad (15)$$

The Gaussian envelope is given by $\alpha(\omega_a, \omega_b) = (2\pi\sigma_p^2)^{-1/2} exp\left[-(\omega_a + \omega_b - \omega_p)^2/(2\sigma_p^2)\right]$, which is inherited from the pump pulse centered around $\omega_p$ with $\sigma_p$ bandwidth. The phase matching condition is included in the two-photon state amplitude [16] $\varphi(\omega_a, \omega_b) = \text{sinc}\left\{L_c\left[\left(\omega_a - \frac{\omega_p}{2}\right)(k'_a - k'_p) + \left(\omega_b - \frac{\omega_p}{2}\right)(k'_b - k'_p)\right]\right\}$. Here $L_c$ is the length of the generating nonlinear crystal, $k'_{a,b}$ are the inverse group velocity at half pump frequency $\frac{\omega_p}{2}$ and $k'_p$ is the corresponding velocity at the central frequency $\omega_p$.

There are two limiting cases for this wavefunction. One is mostly used for ultrafast pulses resulting in wide bandwidth pump. In this case a Schmidt decomposition to pulse-modes is useful. The other is narrowband limit in which the width of the pump envelope $\sigma_p^2$ is set to zero. For a narrowband pump pulse, $\alpha(\omega_a, \omega_b) \to \delta(\omega_a + \omega_b - \omega_p)$, and the conjugate temporal profile $\delta T = \sigma_p^{-1}$ is therefore large. The two-photon wavefunction then takes the form,

$$|\psi\rangle_{nb} = \int d\omega \, \text{sinc}\left[\left(\omega - \frac{\omega_p}{2}\right)T_a + \left(\frac{\omega_p}{2} - \omega\right)T_b\right]|\omega\rangle_a|\omega_p - \omega\rangle_b \quad (16)$$

where we have defined the time variables $T_{a/b} = (k'_{a/b} - k'_p)L_c$. By shifting the frequency variables, we obtain

$$|\psi\rangle_{nb} = \int d\omega \, \text{sinc}[\omega T_{ent}]|\omega + \frac{\omega_p}{2}\rangle_a|\frac{\omega_p}{2} - \omega\rangle_b \quad (17)$$

where $T_{ent} = T_a - T_b$ is the entanglement time. Note, that the wavefunction in Eq. (15) does not generally possess exchange symmetry, since the twin state represents the state of the field modes (amplitudes), which can be distinguished by e.g. group velocity dispersion, polarization etc. In the same time Eq. (16) yields a simplified expression originating from the narrowband pump pulse, where overlapping modes result in the exchange symmetry of the wavefunction typical for the HOM experiment. A more general treatment of the photon wave functions is discussed in [26].

**Data and materials availability** the authors declare that full theoretical details are available in the Supplementary Information. All raw data that support the findings in this study are available from the corresponding authors upon request.

**Acknowledgments**

S.M gratefully acknowledges the support of the National Science Foundation through Grant No. CHE-1953045. K.E.D. acknowledges the support from National Science Foundation of China (No. 11934011), Zijiang Endowed Young Scholar Fund, East China Normal University, and Overseas Expertise Introduction Project for Discipline Innovation (111 Project, B12024).

**Author contributions:** K.D., S.A., B.G., and S.M. contributed to the design and writing of the paper. B.G performed the simulations

**Competing interests:** The authors declare no competing interests.

**Additional information**

**Supplementary information** is available for this paper at

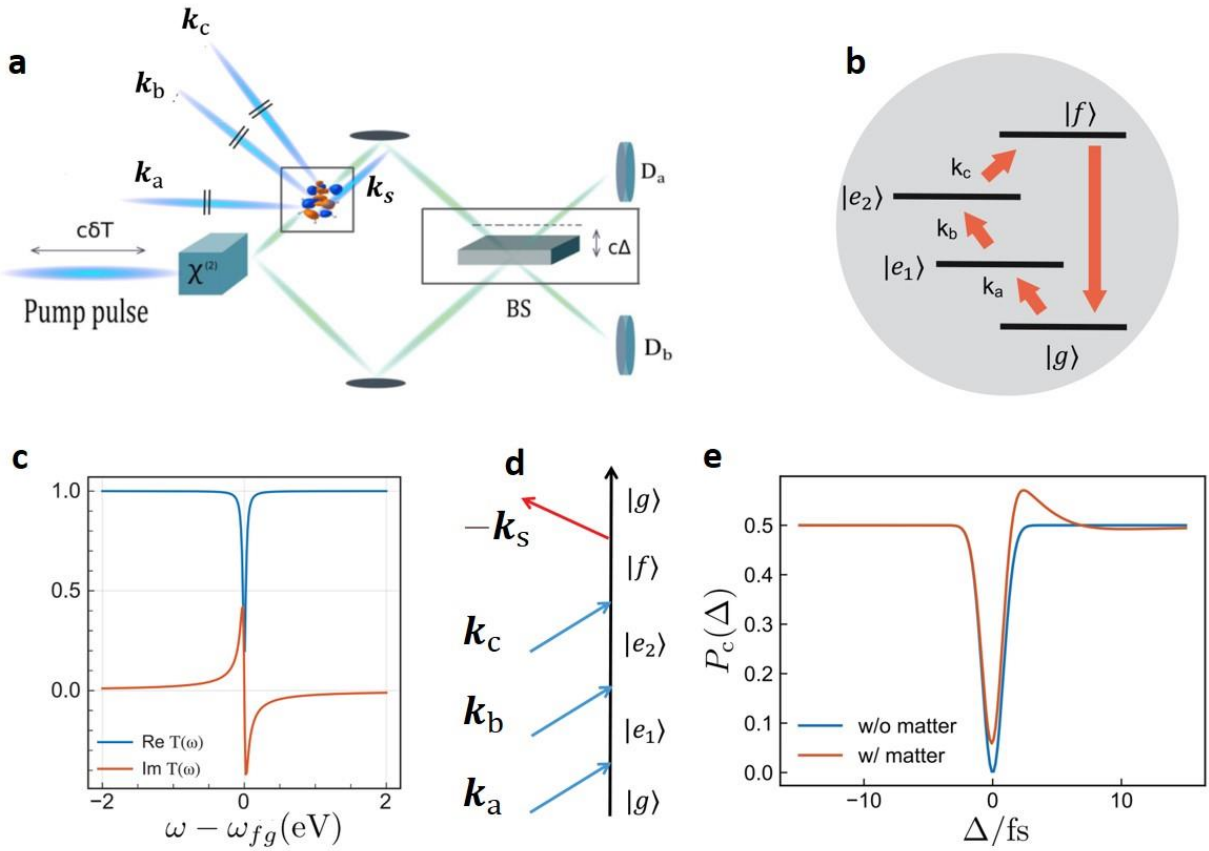

**Fig. 1. Interferometric spectroscopy setup.** Hong-Ou-Mandel interferometer. A pump beam propagates through a $\chi^{(2)}$ nonlinear crystal. The interaction between field modes mediated through the crystal induces entanglement between two well separated beams of different polarization produced by spontaneous parametric down conversion. One beam interacts with a sample (inset I) while the other propagates in an empty arm. The two beams are finally combined on a beam-splitter (BS) (inset II) and collected in two detectors $D_a$ and $D_b$. Insets I and II are specified in more detail in Figs. 3 and 4 respectively.

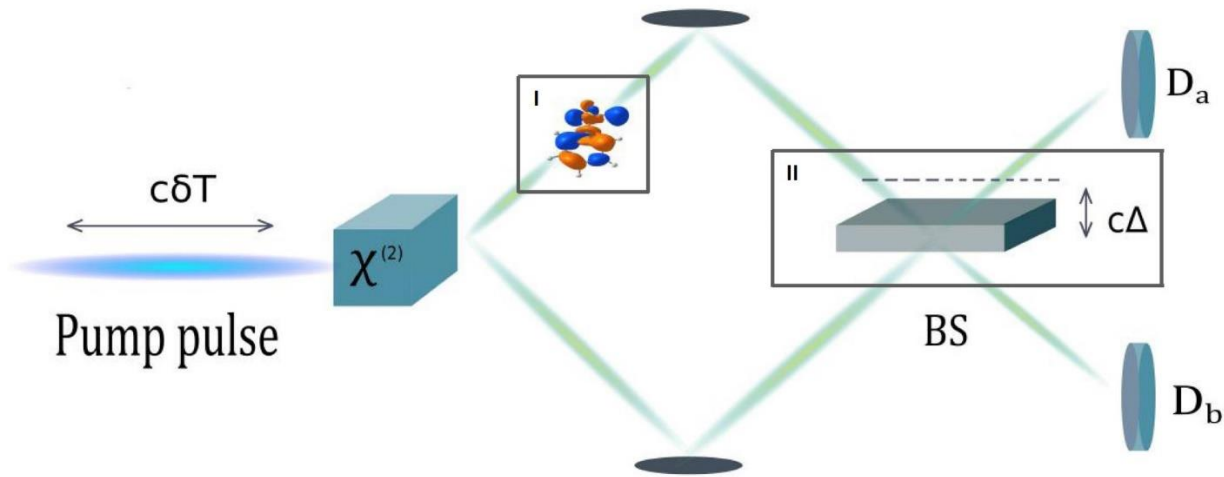

**Fig. 2 Hong-Ou-Mandel (HOM) interference signal.** An illustration of the variation of the coincidence probability with the optical delay $\Delta$. (a) Level scheme of the model system. (b) Experimental setup based on HOM interferometer combined with four-wave mixing signal. (c) Its transmission function $T(\omega) \equiv 1 - iA_0\chi^{(3)}(\omega)$ (real part – orange line, imaginary part – blue line) vs the scanned frequency $\omega_3$ at fixed $\omega_1$ $and$ $\omega_2$. The third-order susceptibility for a multi-level system is computed following Ref. [5]. (d) Schematic diagram representing the main contribution of the third-order susceptibility. (e) Variation of the HOM coincidence counting rates with the optical delay (without matter – blue line, with matter – orange line).

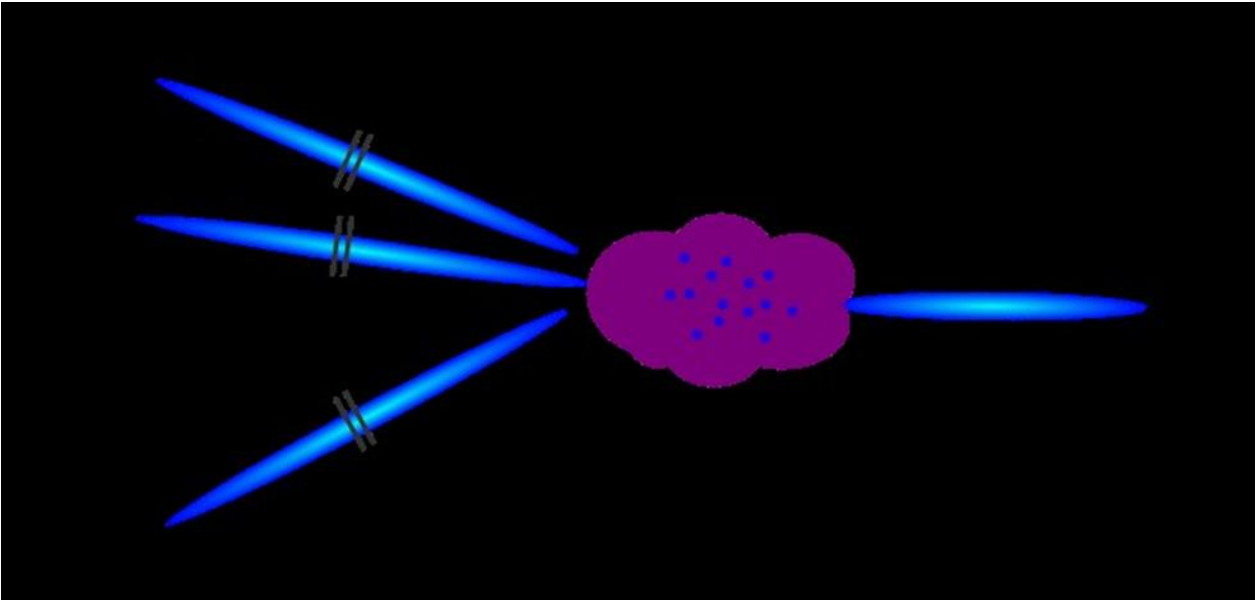

**Fig. 3 Four-wave mixing scheme.** Inset I in Fig. 1 is shown for a four-wave mixing process. Three incoming waves $\boldsymbol{k}_1, \boldsymbol{k}_2, \boldsymbol{k}_3$ interact with matter. The fourth $\boldsymbol{k}_s$ mode is the detected signal in the direction dictated by the phase matching factor $f(\Delta k)$ introduced in Eq. (2).

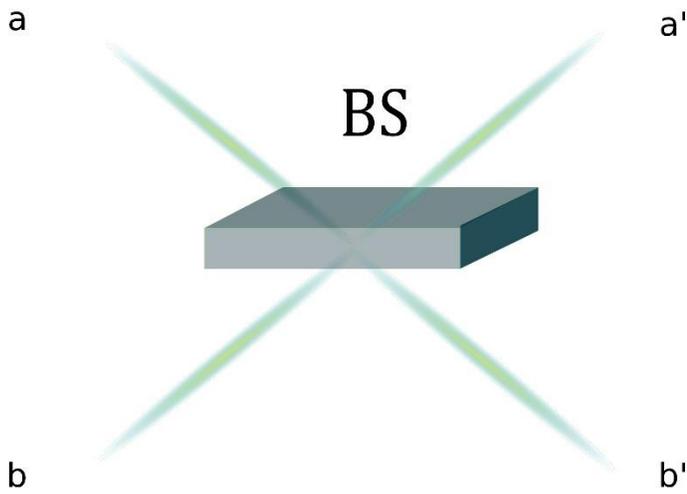

**Fig. 4 The beam splitter.** BS of inset II of Fig. 1 is shown with more details. The BS described in Eq. (4), generates a superposition of the incoming fields (plane subscripts) in its output (primed subscripts).

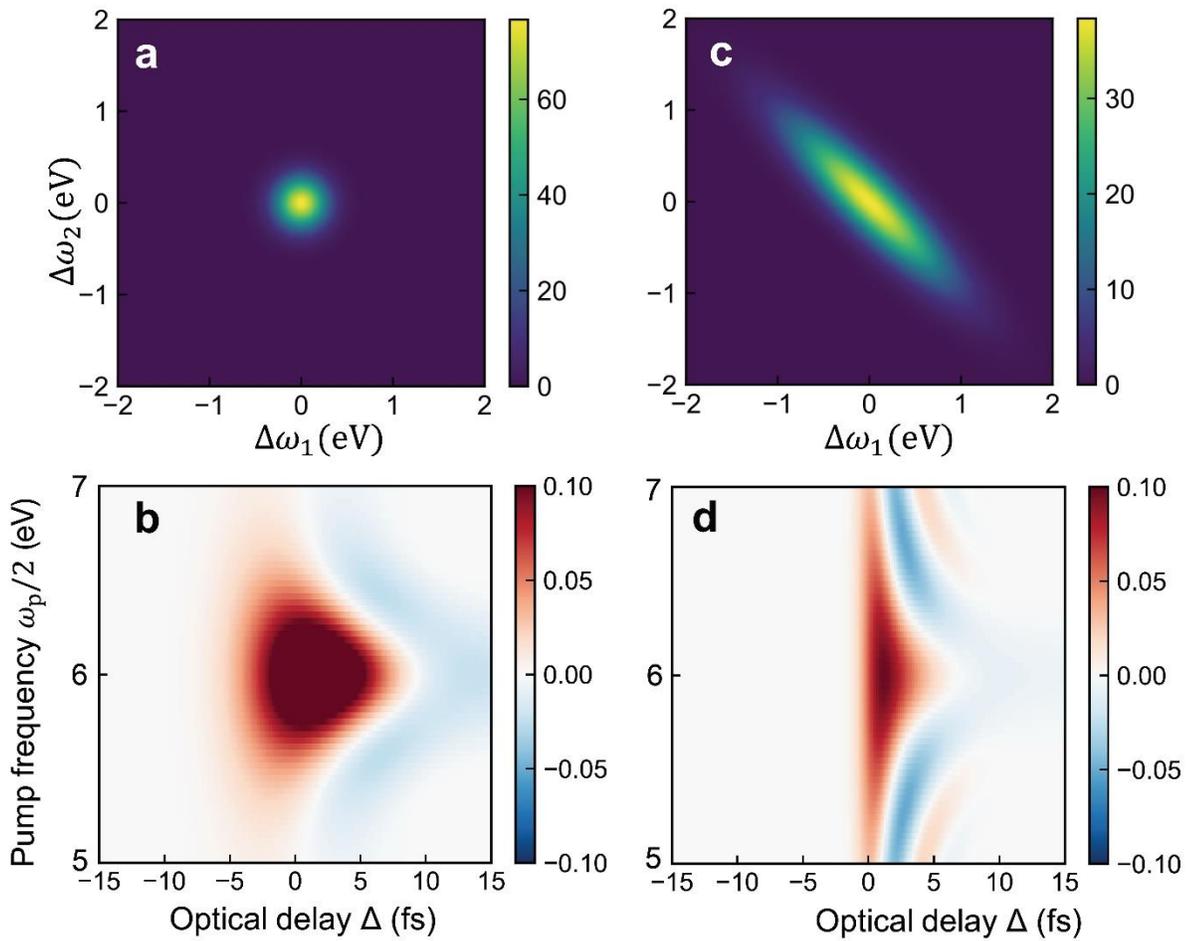

**Fig. 5. 2D photon coincidence counting signal.** $P_{ab}(\Delta, \omega_p)$ [Eq. (7)] for the model system of Fig. 2b. The pump frequency $\omega_p$ creating the entangled photon pair and the optical delay between two arms are varied. We approximate a sinc-shape of the two-photon wavefunction by a Gaussian for simplified frequency integration: $\Phi(\omega_1, \omega_2) = (2\pi\sigma_+\sigma_-)^{-1/2} \exp\left[-(\omega_1 + \omega_2 - \omega_p)^2 / (16\sigma_+^2)\right] \exp[-(\omega_1 - \omega_2)^2/(4\sigma_-^2)]$. (a) Two-photon wavefunction of unentangled photons with $\sigma_+ = 0.1$ eV, $\sigma_- = 0.2$ eV. (b) The Hong-Ou-Mandel (HOM) spectrum using the two-photon state in (a). (c) Two-photon wavefunction of highly entangled twin-photons with energy anti-correlation. Here $\sigma_+ = 0.1$ eV, $\sigma_- = 0.8$ eV. (d) The HOM spectrum using the twin-photon state in (c). The resolution is clearly enhanced by the entangled photon pair.

**Supplementary Information for**

**"Hong-Ou-Mandel interferometry and spectroscopy using entangled photons"**

**Authors**

Konstantin E. Dorfman,[1]* Shahaf Asban,[2] Bing Gu[2], and Shaul Mukamel[2]

1. State Key Laboratory of Precision Spectroscopy, East China Normal University, Shanghai 200062, China

*Email: dorfmank@lps.ecnu.edu.cn

2. Department of Chemistry and Physics & Astronomy, University of California, Irvine, California 92697-2025, USA

**Supplementary Note 1. Interaction picture equation of motion for the electromagnetic field**

In order to introduce the moments of the field obtained by the interferometer through coincidence photon counting, we first derive an expression for the field operator immediately after the interaction with the matter. The interaction picture equation of motion is given by,

$$\frac{d}{dt}\hat{\mathbf{E}}_s^{(+)}(\mathbf{r},\mathbf{t}) = -\frac{i}{\hbar}\left[\hat{\mathbf{E}}_s^{(+)}(\mathbf{r},\mathbf{t}), \hat{H}_{\mu\phi}(\mathbf{r},\mathbf{t})\right]. \quad (S1)$$

Such commutators essentially are diverging and can be regularized by infinitesimal temporal integration. In order to avoid such operation we develop a alter procedure by representing the Heisenberg equation formally as the following limit,

$$\frac{d}{dt}\hat{\mathbf{E}}_s^{(+)}(\mathbf{r},\mathbf{t}) = -\frac{i}{\hbar}\lim_{t'\to t}\left[\hat{\mathbf{E}}_s^{(+)}(\mathbf{r},\mathbf{t}), \hat{H}_{\mu\phi}(\mathbf{r},\mathbf{t}')\right], \quad (S2)$$

The field operator is given by Eq. (10). Eq. (S2) then yields,

$$\begin{aligned}\frac{d}{dt}\hat{\mathbf{E}}_s^{(+)}(\mathbf{r},\boldsymbol{\tau}) &= -\frac{i}{\hbar}\lim_{t'\to t}\left[\hat{\mathbf{E}}_s^{(+)}(\mathbf{r},\mathbf{t}), \hat{H}_{\mu\phi}(\mathbf{r},\mathbf{t}')\right] \\ &= -i\lim_{t_\varepsilon\to 0}\sum_{s'}\frac{\sqrt{\omega_s\omega_n}}{2\pi}e^{-i\mathbf{r}(\mathbf{k}_s-\mathbf{k}_n)}\delta_{ss'}\int d\omega\, e^{-i\omega(t_\varepsilon)}\hat{V}(\mathbf{r},t'),\end{aligned} \quad (S3)$$

where $t_\varepsilon = t - t'$. Integrating this expression and using $\int d\omega\, e^{-i\omega t} = 2\pi\delta(t)$ and Heaviside theta function $u(\tau - t') = \int_{-\infty}^{\tau} dt\, \delta(t - t')$, the field operator prior to the interaction takes the form,

$$\hat{E}_s^{(+)}(\tau) = \hat{E}_s^{(+)}(-\infty) - i\omega_s \hat{V}(\mathbf{r}, t) u(\tau - t). \text{ (S4)}$$

After integration we finally obtain the field operator,

$$\hat{E}_s^{(+)}(\mathbf{t}) = \hat{E}_s^{(+)}(-\infty) - i\omega_s \hat{V}(\mathbf{r}, t). \text{ (S5)}$$

One can further neglect the vacuum background contributions $\hat{E}(-\infty)$.

**Supplementary Note 2. $L^* \gg L_p$ - generalized susceptibilities**

Nonlinear spectroscopy with quantum states of light can be roughly divided into two levels. On the first level which was considered so far, the matter degree of freedom is characterized by a set of causal response functions (susceptibilities) which are the result of consecutive interactions followed by a single measurement [4,11]. In this level the observable is *classical*, and it is measured with a unique probe that in some cases allows higher accuracy due to quantum enhancements [12]. Ultrashort pulses can exploit the quantum nature of light in order to increase the temporal resolution of the *classical response* measurements. The second level contains novel observables which have no classical counterpart. They can be described by a classical response with quantum *noncausal* contributions. In this regime one employs longer pulses which can scramble time ordering and give rise to *generalized quantum response*.

When the pulse duration and the matter response time exceed the free propagation time throughout the interferometer $\{\delta T, \tau_R\} \gg L_p/c$, it is not possible to trace out the matter degrees of freedom prior to the measurement time. For simplicity, in this section we will not discuss in detail specific geometrical characteristics of the sample, and consider a single molecule and focus on the time ordering which gives rise to intriguing effects. Computing Eq. (5) using Eq. (1) transformed by Eq. (4), considering the narrowband wavefunction given in Eq. (14) yields the two photon coincidence probability,

$$P_{ab}(t, \tau, \Delta) = \mathcal{N}^{-1} \hbar \omega_0^3 \{T^2 \text{tr}\left[T\hat{V}_R^\dagger(t+\tau)\hat{V}_L(t+\tau)U(t+\tau)\rho_{\mu\phi_a}(0; t, t)\right]$$

$$+ R^2 \text{tr}\left[T\hat{V}_R^\dagger(t+\Delta)\hat{V}_L(t+\Delta)U(t+\Delta)\rho_{\mu\phi_a}(0; t+\tau-\Delta, t+\tau-\Delta)\right]$$

$$-RT\text{tr}\left[T\hat{V}_R^\dagger(t+\tau)\hat{V}_L(t+\Delta)U(t^*)\rho_{\mu\phi_a}(0;t+\tau-\Delta,t)+h.c.\right] \quad (S6)$$

Here T is the Liouville space time ordering operator, $U(t) = \exp\left[-\frac{i}{\hbar}\int_{-\infty}^t dt' H_{\mu\phi,-}(t')\right]$ is the time evolution operator and $t^* = \max\{t+\tau, t+\Delta\}$. Double excitations proportional to $\propto RT\langle V^\dagger V^\dagger VV\rangle$ are neglected. The density operator $\rho_{\mu\phi_a}(t_1;t_2,t_3) = \rho_\mu(t_1) \otimes \rho_{\phi_a}(t_2,t_3)$ where the (reduced) single photon density matrix after tracing the $'b'$ photon out is given by,

$$\rho_{\phi_a}(t_1,t_2) = e^{-i\omega_p t_1}|\phi(t_1)\rangle\langle\phi(t_2)|e^{i\omega_p t_2}, \quad (S7a)$$

$$|\phi(t)\rangle = \frac{1}{\sqrt{2\pi}}\int d\omega\, \phi(\omega,\omega_p-\omega)e^{i\omega t}\hat{a}^\dagger(\omega)|0_a\rangle \quad (S7b)$$

The first two terms of Eq. (S6) are invariant to the exchange $\tau \leftrightarrow \Delta$. They hold information that can be obtained by photon counting prior to the interference of the upper branch, which can be used in order to eliminate their contribution when desired. The key feature of this expression resulting from the time ordering operator, is the last term of Eq. (S6). The dipole operators are taken at two different times separated by a control parameter $\Delta$. Furthermore, it is clear from Eqs. (S7a), (S7b) that the field coherence plays a significant role. One can $O_\pm A = OA \pm AO$ and the last line of Eq. (S6) are recast as,

$$\langle T\hat{V}_R^\dagger(t_1)\hat{V}_L(t_2)\exp\left[-\frac{i}{\hbar}\int_{-\infty}^{t^*} dt' H_{\mu\phi,-}(t')\right]\rangle_{\phi_a} = \frac{1}{4}\langle T[\hat{V}_+^\dagger(t_1)\hat{V}_+(t_2) +$$
$$\hat{V}_+^\dagger(t_1)\hat{V}_-(t_2)]\exp\left[-\frac{i}{\hbar}\int_{-\infty}^{t^*} dt' H_{\mu\phi,-}(t')\right]\rangle_{\phi_a} \quad (S8)$$

assuming $t_1 > t_2$ and $\langle\cdots\rangle_{\phi_a}$ is done with respect to the reduced density matrix. The term with $V_-$ acting from the left vanishes since the trace of the commutator is zero. In this case, to lowest order when interactions occur during the interval $t_1 < t' < t_2$, observables such as $\langle V_+V_+V_-V_-\rangle$, $\langle V_+V_-V_-V_+\rangle$ and $\langle V_+V_-V_+V_-\rangle$ denoted as generalized susceptibilities contribute to the coincidence signal. These contributions have no classical analogue and correspond to moments of molecular quantum fluctuations induced by the field, where $(+)$ stands for a fluctuation, and $(-)$ an interaction. Note that this signal bears a strong resemblance to the spontaneous incoherent signal studied in [4,11] with one important difference, the time difference between measured interactions is entirely controlled by the experimentalist. In contrast, spontaneous signals require integration over the interaction time.

This signal scrambles the time ordering and gives rise to novel quantum observables, thus providing novel matter observables denoted as generalized susceptibilities. These go beyond the present study.

**Supplementary Note 3. The broadband HOM spectrometer**

When a broadband pump is used, performing Schmidt decomposition to the time-energy entangled photons is particularly useful.

**Schmidt decomposition of broadband pump**

In the case of an ultrashort pump, it is convenient to present the photon-pair using the Schmidt decomposition, whereby the amplitude takes the form [13,14],

$$\Phi(\omega_a, \omega_b) = \sum_n \sqrt{\lambda_n}\, \psi_n(\omega_a)\phi_n(\omega_b). \quad (S9)$$

The single-photon amplitudes obey the coupled integral equations [9],

$$\begin{aligned} \psi_n(\omega) &= \frac{1}{\lambda_n} \int d\omega'\, \kappa_a(\omega, \omega')\psi_n(\omega'), \\ \phi_n(\omega) &= \frac{1}{\lambda_n} \int d\omega'\, \kappa_b(\omega, \omega')\phi_n(\omega'), \end{aligned} \quad (S10)$$

where the kernels are given by tracing over the counter photon,

$$\begin{aligned} \kappa_a(\omega, \omega') &= \int d\omega_b\, \Phi(\omega, \omega_b)\Phi^*(\omega', \omega_b) \\ \kappa_b(\omega, \omega') &= \int d\omega_a\, \Phi(\omega_a, \omega)\Phi^*(\omega_a, \omega') \end{aligned} \quad (S11)$$

and can be interpreted as a single photon spectral correlation. We further define Schmidt temporal-modes $u_n(t) = \frac{1}{\sqrt{2\pi}} \int d\omega\, \psi_n(\omega) e^{-i\omega t}$ and $v_n(t) = \frac{1}{\sqrt{2\pi}} \int d\omega\, \phi_n(\omega) e^{-i\omega t}$ which will be used below.

**The broadband HOM signal**

We now examine the ultrafast variant of this experiment. Using Eq. (13) followed by Schmidt decomposition defined in Eqs. (S9) – (S11) and the pulse-modes we obtain,

$$P_{ab}(\tau, \Delta; \{\omega\}_n) = P_0 \left\{ T^2 \left|\sum_k \sqrt{\lambda_k}\, v_k(t)\tilde{u}_k(t+\tau)\right|^2 + R^2 \left|\sum_k \sqrt{\lambda_k}\, v_k(t+\tau-\Delta)\tilde{u}_k(t+\Delta)\right|^2 - \right.$$

$$\left. RT\left[\left(\sum_k \sqrt{\lambda_k}\, v_k^*(t)\tilde{u}_k^*(t+\tau)\right)\left(\sum_m \sqrt{\lambda_m}\, v_m(t+\tau-\Delta)\tilde{u}_m(t+\Delta)\right) + c.c.\right] \right\} (S12)$$

where $\tilde{u}_k(t) = u_k(t) * \chi^{(n)}(t)$. The coincidence probability clearly reflects that only pulse-modes of the upper branch ($u_k$) are modulated by the matter response function. The noninteracting pulse modes ($v_k$) are therefore not modulated by the response function and can be used to employ mode selection techniques. This fact can be useful when single modes are detectable to study the response function one mode at a time. This cab be employed by quantum state tomography [3] after the interaction with the matter. One benefit from this procedure is improved frequency resolution, similar to the protocol demonstrated in [15] for spatial modes. By reweighting the modal contributions of the combined signal, more dense frequency information can be revealed along the lines of the explanation below.

**Post-measurement pulse-shaping**

When the twin state is fully characterized (in the absence of matter), complete knowledge of the Schmidt weights $\lambda_n$ can be achieved. Using the completeness and closure relations for the basis sets $\{\psi_n\}$ this knowledge may be used for post measurement pulse-shaping. The completeness of $\psi_n$ reads,

$$\delta(\omega - \omega') = \sum_n \psi_n(\omega)\psi_n^*(\omega'). \text{ (S13)}$$

Suppose one is interested in a pulse envelope given by $A(\omega)$ which is not achievable experimentally. Since the weights of each mode is known, a simple reweighting can achieve a desired pulse envelope. Multiplying Eq. (S13) from the left and right by $A(\omega')$ and integrating w.r.t. $\omega'$ yields $A(\omega) = \sum_n a_n \psi_n(\omega)$ where $a_n = \int d\omega'\, \psi_n^*(\omega')A(\omega')$. Summing the post-measurement results mode by mode with the corresponding weights results in the desired $A(\omega)$. One example of such resummation is by assuming the weights $a_n = \psi_n^*(\omega_s)$ which can be used to scan the signal using a tunable narrowband profile with variable frequency $\omega_s$. When infinite number of modes are available, the pulse envelope converges to delta distribution $A(\omega) \to \delta(\omega - \omega_s)$.

**Supplementary Note 4. Mach-Zehnder Interferometric Spectroscopy**

We now consider the setup of Fig. S1 in which the incoming photons (not necessarily entangled in this scenario) first interfere, here both photons interact with the matter. We assume that only the second BS in Fig. S1 in translated by $\pm\Delta$, posterior to the interaction with the matter. It is straightforward to generalize the resulting expressions to include additional shift in the first BS. We monitor photon counting difference as a function of the shift parameter $\Delta$. The input-output relation for the field operators after the first and second BS is given with a single- and double prime notations respectively,

$$\begin{aligned}\hat{\mathbf{E}}_{a''}^{(+)}(t) &= \sqrt{T}\hat{\mathbf{E}}_{a'}^{(+)}(t) - i\sqrt{R}\hat{\mathbf{E}}_{b'}^{(+)}(t+\Delta),\\ \hat{\mathbf{E}}_{b''}^{(+)}(t) &= \sqrt{T}\hat{\mathbf{E}}_{b'}^{(+)}(t) - i\sqrt{R}\hat{\mathbf{E}}_{a'}^{(+)}(t-\Delta),\end{aligned} \quad \text{(S14)}$$

immediately after the second BS. When both beam splitters have the same transmission coefficients, The field operator prior to the interaction with matter is given in terms of the incoming field by $\hat{\mathbf{E}}_{a'}^{(+)}(t) = \sqrt{T}\hat{\mathbf{E}}_{a}^{(+)}(t) + i\sqrt{R}\hat{\mathbf{E}}_{b}^{(+)}(t)$. Note that we place the beam-splitters facing the opposite branch to cancel unwanted added relative phases according to the Fresnel convention, hence the relative minus sign in the reflection coefficient.

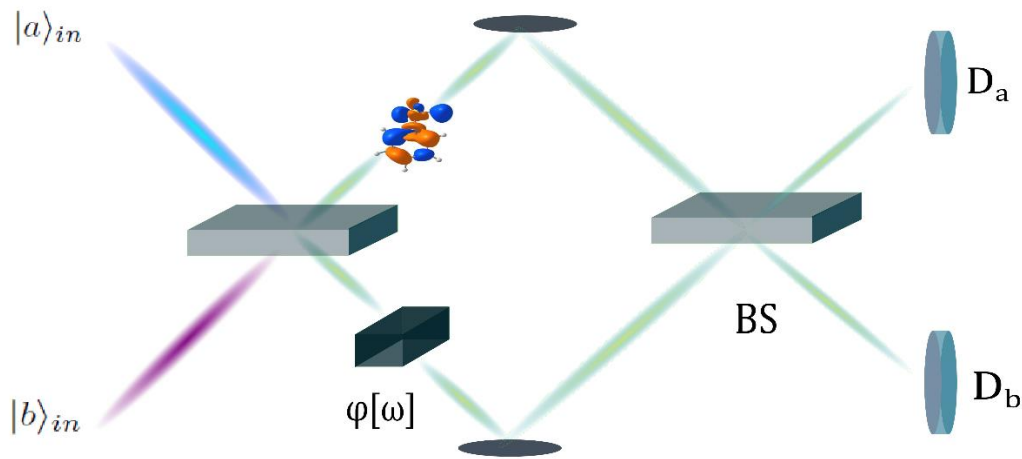

**Fig. S1 Mach-Zehnder interferometric spectroscopy setup.** Two incoming photons are mixed at a beam splitter. One arm contains the molecular sample, while the other passes through a phase element $\varphi[\omega]$. After mixing both arms on a beam splitter the resulting coincidence detection between detectors $D_a$ and $D_b$ is measured.

We obtain the number difference between the two ports $\langle \hat{n}_{b''}(t+\Delta) - \hat{n}_{a''}(t) \rangle$ for a balanced BS ($T = R = \frac{1}{2}$) where $\langle \hat{n}(t) \rangle = \left\langle \hat{\mathbf{E}}_R^{(-)}(t)\hat{\mathbf{E}}_L^{(+)}(t) \right\rangle$. Using Eq. (S14) the photon difference signal is given by,

$$S(t,\Delta) = \langle \hat{n}_{b''}(t+\Delta) - \hat{n}_{a''}(t) \rangle \text{-2Im} \langle T\hat{E}_{a',L}^{(-)}(t)\hat{V}_L(t+\Delta)\exp\left[-\frac{i}{\hbar}\int_{-\infty}^{t^*}dt'\,H_{\mu\phi,-}(t')\right]\rangle \quad \text{(S15)}$$

where $\sqrt{2}H_{\mu\phi}(t) = \hat{\mathbf{E}}_{b'}^{(-)}(t)\hat{V}(t) + \hat{\mathbf{E}}_{b'}^{(+)}(t)\hat{V}^\dagger(t)$ and $\sqrt{2}\hat{\mathbf{E}}_{b'}^{(+)}(t) = \hat{\mathbf{E}}_{b}^{(+)}(t) - i\hat{\mathbf{E}}_{a}^{(+)}(t)$ is the combined field that is coupled to the matter. Eq(S15) bears strong resemblance to the one introduced in [11] for $\Delta = 0$. However, there are two fundamental differences. First, there is a phase difference between the field in the definition of the signal and the one in the coupling Hamiltonian. Second, Eq. (S15) depends on the additional controlled time difference $\Delta$.

In this setup, it would be useful to employ either a multimode squeezed state or entangled pair initial state of the probe. Both are two-photon states and provide a rich optimization playground for quantum enhancement of metrology applications. Eq. (S15) essentially measures phase difference between the two incoming beams. The number-phase uncertainty may be exploited to demonstrate the measurement accuracy as a function of controlled variables such as the squeezing parameters or entanglement time.